# CMOS image sensor with micro-nano holes to improve NIR optical efficiency: micro-holes on top surface vs on bottom

E.Ponizovskaya Devine[1,2] *Member, IEEE*, Ahasan Ahamad[2] *Member, IEEE*, Ahmed Mayet[2] *Member, IEEE*, Amita Rawat[2] *Member, IEEE*, Aly F Elrefaie[1] *Life Fellow, IEEE*, Toshishige Yamada[1] *Senior Member, IEEE*, Shih-Yuan Wang[1] *Life Fellow, IEEE*, M Saif Islam[2] *Fellow, IEEE*

*Abstract*— We study the nano- and micro-structures that increase the optical efficiency of the CMOS image pixels in visible and infrared. We consider the difference between the micro-holes at the pixels' bottom and the top and the holes that are composed of smaller holes. Those solutions can facilitate the fabrication. We study the crosstalk and the optical efficiency dependence on the angle of incident of light and numerical aperture for the pixels in a camera setup.

## I. INTRODUCTION

SEVERAL solutions are proposed addressing the need for the complementary metal oxide semiconductor (CMOS) compatible high sensitivity in infra-red (IR) cameras for smartphones, automobiles, security, robotics, and augmented/virtual reality device applications [1]–[5], particularly, the device solutions involving micro/nano structures for light-trapping [6-11]. Recently, it was shown that one hole per pixel provides better optical efficiency [6]. A micro-structured surface was shown to increase the crosstalk between the pixels and the crosstalk is reduced by the deep trenches between the pixels [12]. The Ge-on-Si pixels with micro-holes are shown to increase the infrared operation wavelength limits to a longer wavelength [8]. One of them is having holes at the bottom of the pixel. Another is using small nanoholes that compose a larger hole of a size optimal for optical efficiency in infrared. For the composite holes, we have chosen a hexagonal nanohole symmetry that possesses a high rotational symmetry. It allows the shaping of the photonic crystal in the composite cylindrical hole. The property of such a composite hole will be determined by the property of the shape of the larger scale hole and the property of the lattice of small holes.

We consider a Si-on-insulator (SOI) substrate with a $1.12 \times 1.12$ μm² pixel size. We show the impact of nano-microstructures at the bottom of the pixel on the optical efficiency using Finite Time Domain (FDTD) [13,14] Lumerical simulations. We have shown that the microstructure on the bottom can produce a similar effect as the nanostructured surface. It includes the fact that the crosstalk increased by the inverted pyramids was reduced using deep trench isolation (DTI) [12,13]. The previous simulations studied only the normal incident. However, the light incidence at an angle could produce higher crosstalk because the angle with the DTI is increased. We also performed the simulation with the realistic numerical aperture (NA) that collects the light at an angle of ±10 degrees exchange between neighboring pixels (crosstalk). We have shown that the DTI keeps the crosstalk in the range comparable with flat devices without nanostructured pixels.

## II. DESIGN AND OPTICAL SIMULATION

In simulations, we consider Si backside-illuminated sensors layered over a signal processing circuit chip with a low-noise structure [6], [7]. The optical simulations do not include metal contacts. The contacts will add small optical losses that can reduce the results by a few percent. The pixel array is shown schematically in Fig. 1. Each pixel is 1.12 μm wide and consists of a Si layer on $SiO_2$. The Si thickness is 3 μm.

The buried $SiO_2$ layer reduces the transmittance from the thin absorbing layer by reflecting the illuminated electromagnetic waves. A micro-lens of radius 1 μm is used at the top of each pixel, along with a filter. The micro-lens refractive index is 1.5, and the thickness of the micro-lens is 500nm, which is a typical parameter for image sensor design. We model pixels in a Bayer array (Fig.1c) with commonly used color filter parameters [14]. Transmittance after the filters is shown in Fig. 1b. In the 3D simulation setup, the transmittance associated with 900nm thickness is used for the pigment filters [14]. The view of the cross-section through the pyramids is shown in Fig.1d.

We calculate the optical efficiency (OE), which represents the absorption of light by the active layer of each pixel as it was described in [6] and is the difference between the real parts of the Poynting flux entering the volume ($P_{in} = \text{Re}(E \times H^*)$),

[1] W&WSens Device Inc
[2]. University of California, Davis
Contact email: eponizovskayadevine@ucdavis.edu

and the Poynting flux leaving the pixel ($P_{out}$), that is simulated with FDTD method:

$$OE = \frac{P_{in} - P_{out}}{P_{inc}} \quad (2)$$

where $P_{inc}$ is the Poynting vector of the incident light calculated above the lenses and filters. In this study, it was assumed that the quantum efficiency is proportional to the optical efficiency [6].

The quantum efficiency (QE) on the other hand is the photo-current of the device normalized to the light intensity and is less than the OE due to recombination and other losses. Still, the higher OE translates to a higher QE, and optimizing the light-trapping quality would increase QE. The maximum optical efficiency for the blue, red, and green filters determined by the filter transmittance is $OE_{Blue}$ = 75% at 440 nm, $OE_{Green}$ = 80% at 550 nm, and $OE_{Red}$ = 80% at 650 nm, the filters are transparent at the near-infrared region (Fig. 1b). The Si optical absorption is very weak at infrared and light-trapping strategies are required to enhance the absorption, leading to higher optical efficiency.

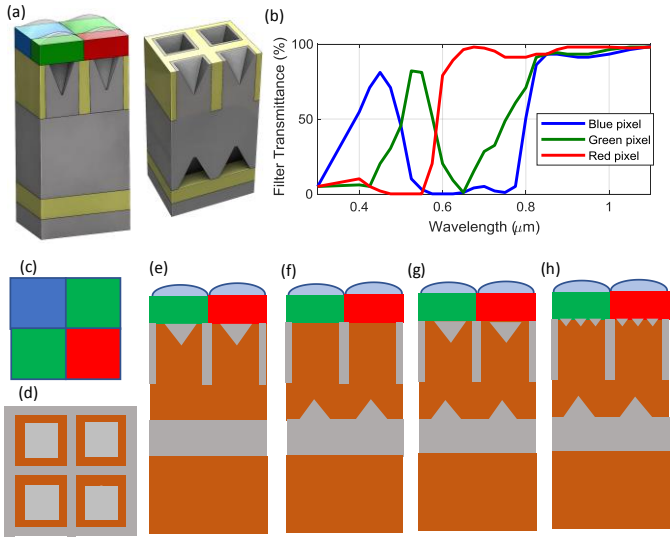

Fig. 1. The schematics of the pixel: a) 3D view of the image sensor with lenses and filters on the right and the middle part with holes on the top and the bottom in detail, b) filters transmittance c) a top view of the Bayer pattern, d) cross-section through the holes pyramids at the bottom, e) side view for holes pyramids on the top, f) side view for holes pyramids on the bottom and bottom g) side view for holes on top and bottom h) side view of the pixel with single pyramid hole 900×900nm at the bottom and an array of small pyramids at the top.

To increase the OE, we incorporate a pyramid hole at the center of each pixel as shown in the side-views Fig.1 (d-h). The microholes are positioned at the top just behind the filter (Fig.1e), at the bottom above the SiO$_2$ layer (Fig.1f), and at the top and bottom (Fig.1g). The side size of the hole is 900nm that showed good quantum efficiency for photodetectors in infrared. The design with a single pyramid at the bottom and an array of small pyramids at the top was also considered (Fig.1h). The array at the top consisted of pyramids of 200×200nm$^2$ with a 250nm distance between the pyramids' centers. There are trenches at the edges of each pixel with a width of 150 nm and a depth of 2μm to reduce the crosstalk. The depth of the pyramid holes is determined by the fabrication process that keeps the angle between the sidewall and the bottom surface is 54.74° which is the angle between the (111) plane and the (100) plane of monocrystalline Si. Micro-hole arrays were shown to enhance the optical absorption of Si for the 800-1000nm wavelength range [9,6] and for the Ge, the wavelength range can be increased up to 1700nm [8]. The study [6], [7] shows that a single hole per pixel provides a better increase in quantum efficiency. The optimal microhole size is comparable to the small-sized pixels. The deep trench structure with the Si-SiO$_2$ interface is used as a barrier against electron diffusion and reflects light to prevent crosstalk. The micro-structure redirects the normal incident light into lateral directions parallel to the surface plane and helps increase the absorption and reduce reflection. We used the FDTD method provided by the Lumerical software to solve Maxwell's curl equations numerically for the unit cell that consists of the Bayer array (Fig. 1). We use Bloch boundary conditions in the XY plane and Perfectly Matched Layer (PML) in the direction normal (Z) perpendicular to the surface.

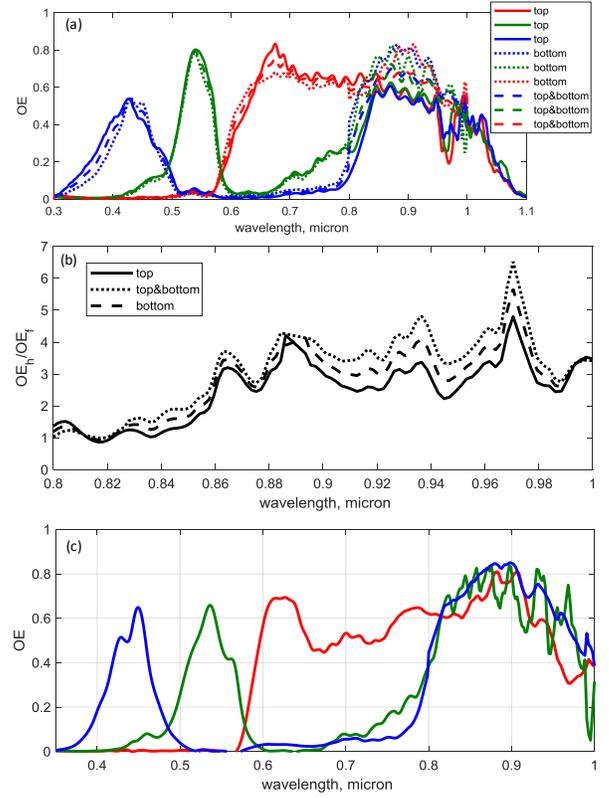

Fig. 2. a) optical efficiency for pyramids at the top, at the bottom, and both b) increase of OE due to micro/nano-holes in the infrared wavelengths, the OE is calculated as the averages of blue, green and red pixels c) OE with pyramids 900×900nm on the bottom and an array of small pyramids 200×200nm on the top with distance 250nm between the centers.

The simulations for the nano-micro structured sensors were compared to the simulation for flat image sensors for vertical illumination. The results for pyramid holes with side 900nm on the top and on the bottom of the pixel as well as on the top and the bottom are presented in Fig.2a.

The results showed that the pyramids at the top can be replaced by pyramids at the bottom with the same or even better effect for some wavelengths. Fig.2b shows the ratio between the OE for the pixels with holes and flat pixels that represent the enhancement that was achieved due to the micro-hole structure. The enhancement is higher for the lower absorption of the pixel material. The OE represents the average enhancement over the red, green, and blue pixels for Fig.2b as all filters' transmittance is almost the same for wavelengths longer than 800nm.

The simulations show that the OE can be improved 3 times for infrared. The mechanism of the light trapping is similar to the holes on the top. The light is dispersed by the micro-hole in the pixel. As it was shown in [9] we can see the holes produce the lateral modes that are trapped in the pixel. The difference is the microstructured surface has less reflection. This effect was compensated by applying antireflection coating at the top of Si. The holes on both top and the bottom also can serve the same purpose. Small pyramid arrays at the top can reduce the reflection. We study the possibility to combine the small array of holes at the top and a single pyramid 900×900nm² at the bottom. The results are shown in Fig.2c for an array of small pyramids at the top with a size of 200×200nm² and a distance of 250nm between them in a square lattice. The OE in infrared shows results is as good as for the single pyramids at the top and the bottom. However, it is slightly smaller in green and red regions. The reason is that the antireflection coating and the small pyramid array are optimized for different wavelengths. For the same reason, the OE in blue is slightly higher for the small pyramid array than for the single hole on the top. Despite promising OE enhancement with the holes at the bottom, we do acknowledge the added device fabrication complexities with it. However, with the advancement of various wafer bonding methods, SmartCut$^{TM}$ process including the 3D integration, the proposed *holes at the bottom* devices can be conveniently fabricated.

### III. COMPOSITE HOLES

To facilitate the fabrication, we can use composite holes instead of single micro-hole. We are replacing the single hole with a cluster of nanoholes (Fig.3a-c) filled with SiO2. As we have seen before the optimal hole that produced the best effect is a micro-hole with a side of 800-900nm. For the long wavelength, we can consider small holes to satisfy the conditions for Maxwell-Garnett approximation for the effective dielectric permittivity of the composite material. In the approximation of wavelength much longer than the size of the nanoholes the dielectric permittivity is

$$\epsilon = \epsilon_{Si} f + \epsilon_{SiO_2}(1-f)$$

where f is the fraction of Si. As the contrast of the refractive index of composite holes to the refractive index of Si is lower than for $SiO_2$ fillers we expect that the effect for composite holes is smaller. We model the composite holes as a cluster of nanoholes with $SiO_2$ filler. We vary the hole size, the distance between the holes and the holes depth.

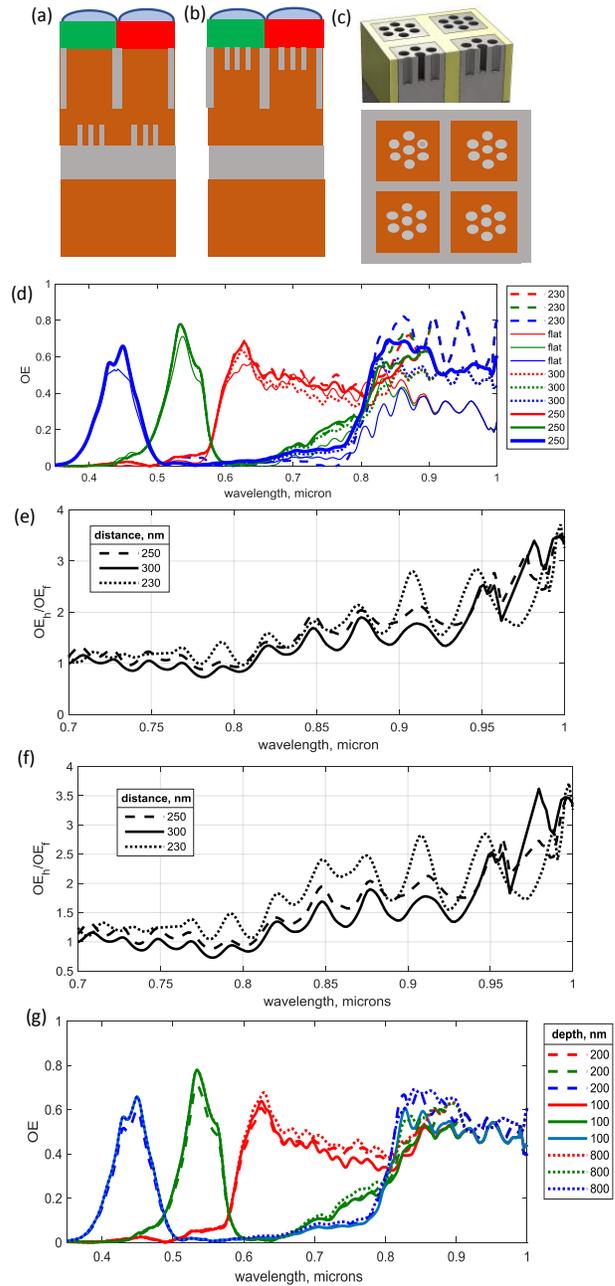

Fig. 3. a) side view of the composite hole on top, b)side view of the composite hole on the bottom, c) 3D image of the composite holes and the cross-section through the composite holes filled SiO$_2$ d) optical efficiency for composite holes on the bottom of several sizes: dashed d=200nm a=250nm, solid d=200nm a=300nm, dotted d=200nm a=230nm the depth is 200nm (e) OE enhancement in infrared for the composite hole on the bottom with nanoholes depth 200nm, diameter 200nm and distance between the nanoholes 230nm (dotted), 250nm(dashed) and 300nm (solid). (f) OE enhancement in infrared for the composite hole on the top with nanoholes depth 200nm, diameter 200nm and distance between the nanoholes 230nm (dotted), 250nm(dashed) and 300nm (solid). g) the composite holes for d=200nm, a=250nm, with depth varying from 100nm to 800nm

The nano-holes arranged in the lattice can support resonance wavelengths. This effect could be seen in photonic crystals. The smaller nanoholes lattice produces resonances for wavelengths in the shorter wavelengths. However, due to the fabrication purpose, the study was focused on the nanoholes with diameters close to 200nm. The diameter size was restricted by the fabrication. A lattice disorder destroys those resonances, however, for the 200nm hole diameter and only a few nanoholes in the composite structure, the disorder influence was not very significant. The depth of the nanoholes varied from 50nm to 1μm.

According to the Maxwell-Garnett approximation, the effective dielectric permittivity of the composite hole is varied from 2.1 to 2.7 with a nanohole diameter of 200 nm and the distance between the centers of the holes varies from 230 to 300nm. A higher dielectric permittivity contrast produces better coupling into lateral modes so the 230 nm distance between nanohole centers that corresponds to effective dielectric permittivity 2.1 produce better absorption.

The results of the FDTD simulations with Lumerical are shown in Fig.3. The flat device optical efficiency is shown in Fig.3a with the thin solid curve. The pixel geometry is shown in Fig. 3a-c. The side view with the composite holes in the bottom is shown in Fig.3a, at the top is shown in Fig.3b, the cross-section through the holes is shown in Fig. 3c. Since the size of the pixel is 1.12×1.12 μm² and the nanohole diameter is only at least 200nm the composite hole consists only of 7 nanoholes arranged in hexagonal lattice.

The optical efficiency is shown in Fig.3d. The FDTD simulations show that the distances 230, 250, and 300nm produce very similar optical efficiency enhancement in infrared both for the holes at the top and at the bottom with a higher enhancement for the shorter distances. The enhancement also increases with wavelength as the absorption of the Si is small and the optical efficiency of the flat device is small as well. As we can see from Fig.3d the optical efficiency for flat devices decreases faster than for the nanostructured pixel. We also see a small increase in optical efficiency in the visible range. The composite hole was optimized for maximum enhancement in infrared. However, the hexagonal nanoholes can produce a guided resonance in the visible and produce a good enhancement for the visible range. Comparing the results with Fig.2 for the holes 900×900nm² with SiO2 we can see that the higher enhancement for the composite holes in blue and green but smaller in red and infrared. In the area for red and infrared the Maxwell-Garnett approximation become valid, and the effective permeability is higher than for the SiO2 which leads to smaller enhancement. The size of the solid SiO2 holes is the same as that optimized for the photodetectors [15]. Disorder in the hexagonal lattice can destroy the guided resonance. It can reduce the OE, but it still should be at least as good as for the single pyramid per pixel. The enhancement in infrared for the composite holes at the top and at the bottom is shown in Fig.3e and f. The figure uses the average OE over blue, green, and red pixels.

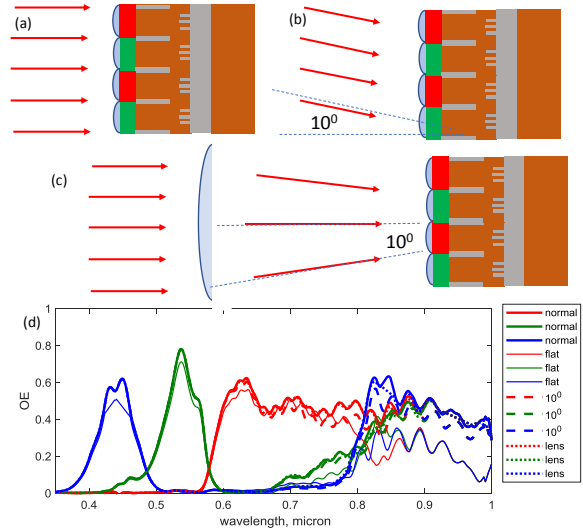

Fig. 4. The schematics of the image sensor with the lens: normal incident light (a), incidence at 10 degrees (b), lens collects light in 10degrees cone (c); Optical efficiency for normal incident light (solid line), compared with light at 10 degrees incidence (dashed line) and light collected by a lens with NA=0.26 (dotted line) Thin line is flat device with lens (d).

The dependence on the depth of the nanoholes was studied (see Fig.3g). It was found that the optical efficiency is not sensitive to the depth with holes deeper than 200nm. At less than 100nm the optical efficiency is close to the one of the flat device, The long holes don't produce significantly higher OE than the composite holes with a depth of about 200nm.

## IV. SIMULATIONS FOR THE LIGHT CONE

In the application configuration such as the camera the pixel array can experience the incident light that is collected within the angle θ, as it is shown in Fig.4. Typically, the light is collected within the angle θ of about 10 degrees. We simulated the lens that collects the light within an angle and compared it to the normal incidence and the plane wave at the angle of 10 degrees. We used the pixels array with a single per-pixel composite hole at the bottom and compared it with the flat device with a lens and without the lens and the normal incident plane wave. The lens that collects light in a cone 10 degrees has a numerical aperture (NA) equal to $n \sin \theta$ and assuming that $n=1,5$ the NA value is *0.26*. The lens was simulated above the pixel array as a spherical surface with n=1.5 and curvature radius to focus length ratio R/f = 0.174. It corresponds to collecting the light at different angles that can increase the crosstalk. Fig.4 shows the angle dependence. Indeed, the crosstalk increases with angle. As the light is not normal it could have a sharper angle with DTI and produce higher crosstalk. We calculated the crosstalk index so that the index in pixel i from pixel j is the integral OE within ±10nm near the wavelength designated for pixel I divided by the same integral OE for the pixel j [6]. The higher value of the index the smaller the crosstalk. The results of the crosstalk are summarized in Table 1.

TABLE I
CROSSTALK CROSSTALK FOR THE NORMAL INCIDENCE COMPARED WITH INCIDENCE AT 10 DEGREES AND LIGHT COLLECTED BY THE LENS WITH NA=0.26

| b/r | b/g | g/r | g/b | r/g | r/b |             |
|-----|-----|-----|-----|-----|-----|-------------|
| 146 | 20  | 56  | 57  | 85  | 34  | Normal, w/h |
| 31  | 14  | 14  | 59  | 36  | 31  | Lens, w/h   |
| 22  | 11  | 11  | 12  | 32  | 18  | 10deg, w/h  |
| 32  | 15  | 15  | 60  | 38  | 32  | Lens, flat  |
| 146 | 24  | 61  | 63  | 89  | 39  | Normal, flat|

As it is expected the crosstalk for the lens is slightly higher than for the normal incident but smaller than for the plane wave of 10 degrees. The flat device and the device with holes and DTI crosstalk is of the same order.

As it was expected, the normal incidence gives the lowers crosstalk. However, the angle of 10 degrees or the lens with a 10-degree cone does not increase the crosstalk beyond the typical values [6].

The OE for a normal incidence (Fig.4a), for a plane wave with an incidence at $10^0$ (Fig.4b), and for the lens that collects light in the cone of 10 degrees (Fig.4c) are shown in Fig.4d, compared to a flat device at 10-degree incidence. The incidence angle slightly effect OE. Still, there is a significant improvement in OE compared with flat devices especially in infrared.

## V. Conclusion

We simulated several configurations of the nano-micro-holes OE enhancement structure for the CMOS image pixel and optimized the size. It was shown that the micro-holes at the bottom of the pixel showed even better enhancement while they have advantages. The single holes also can be replaced with a composite hole that represents a cluster of nanoholes a small degradation in enhancement compared to a single solid hole. Still, it will provide better OE than the uniformly distributed nano-holes.

## Acknowledgment

This work was supported in part by the S. P. Wang and S. Y. Wang Partnership, Los Altos, CA.

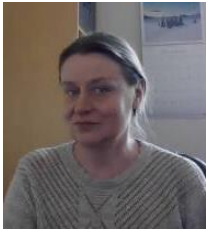
**Ekaterina Ponizovskaya Devine** received the M.S. and Ph.D. degrees from the Moscow Institute of Physics and Technology (State University), Moscow, Russia, in 1999.
She was with the Ames Center, NASA, Mountain View, CA, USA, where she was involved in the optimization and physics-based models for prognostics and automation. She is currently with W&WSens Device, Inc., Los Altos, CA, USA, where she is focusing on photonics and photodetectors.

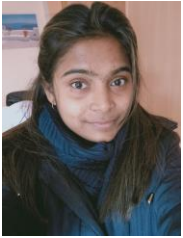
**Amita Rawat** received her B. Tech. degree in Electrical Engineering from IIT Patna, India, and her Ph.D. in Microelectronics (logic device variability modeling and device fabrication) from the Department of Electrical Engineering IIT Bombay. She worked in IMEC Leuven, Belgium on compact modeling of advance logic transistors. She is currently working as a postdoctoral fellow at University of California, Davis, on optoelectronic device modeling, simulation, and fabrication.

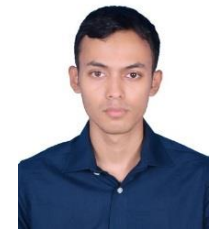
**Ahasan Ahamed** is a second-year graduate student pursuing his Ph.D. degree in the Electrical and Electronics Engineering Department at the University of California, Davis. He received a B. Sc. Degree from Bangladesh University of Engineering and Technology in 2018. He is currently working on spectral response engineering of photon-trapping photodiodes paving towards spectrometer-on-a-chip. His research works also include designing ultra-fast photodiodes and SPADs for hyperspectral imaging.

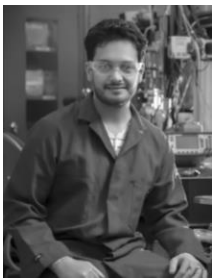
**Ahmed S. (Surrati) Mayet** received the B.Sc. degree in physics from Taibah University, Medina, Saudi Arabia, and the master's degree in electrical and computer engineering from the University of California, Davis, CA, USA, in 2017, where he is currently pursuing the Ph.D. degree with the Electrical and Computer Engineering Department. His research focuses on the development of high-speed and high-efficiency photodetectors for optical communication systems.

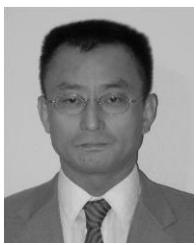
**Toshishige Yamada**( received the B.S. and M.S. degrees in physics from the University of Tokyo, Japan, and the Ph.D. degree in electrical engineering from Arizona State University, AZ. Following a postdoctoral fellow in applied physics at Stanford University, CA, he joined the NASA Ames Research Center, CA and Santa Clara University, CA. He is currently an Adjunct Professor at the UC Santa Cruz, CA and studies the theory and simulations of advanced semiconductor materials and devices.

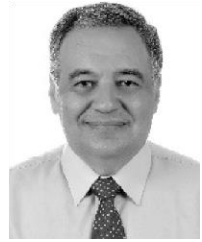
**Aly F. Elrefaie** received (M'83–SM'86–F'04–LF'19) received the B.S.E.E. degree (Hons.) from Ain Shams University, Cairo, Egypt, in 1976, and the M.Sc. and Ph.D. degrees in electrical engineering from NYU, Brooklyn, NY, USA, in 1980 and 1983, respectively.

Since 2014, he has been a Chief Scientist with W&Wsens Devices, Inc., Los Altos, CA, USA, where he is focusing on topics related to nano technology in optical communications

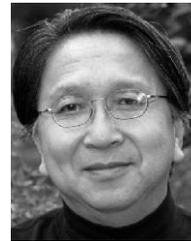
**Shih-Yuan (SY) Wang** B.S. Engineering Physics 1969, UC Berkeley; Ph.D. Electrical. Engineering and Computer Sciences, 1977 UC Berkeley. At HP Labs, SY worked on multimode vertical cavity surface emitting lasers and high speed III-V photodiodes, both of which became successful products. SY is currently with, W&WSens Devices, Inc. Los Altos, CA, working on silicon photodiodes compatible with CMOS process for CMOS image sensors, optical interconnects and LiDAR applications

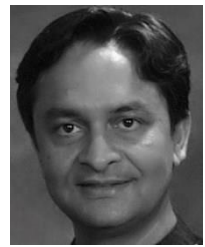
**M. Saif Islam** (SM'15) received the B.Sc. degree in physics from Middle East Technical University, Ankara, Turkey, in 1994, the M.S. degree in physics from Bilkent University, Ankara, in 1996, and the Ph.D. degree in electrical engineering from UCLA, Los Angeles, CA, USA, in 2001.
He joined the University of California, Davis, CA, USA, in 2004, where he is currently a Professor.